\documentclass[aps,prd]{revtex4-2}

\usepackage{eurosym}
\usepackage{amsmath,amssymb}
\usepackage{graphicx}
\usepackage{dcolumn}
\usepackage{bm}
\usepackage{color,cancel}
\usepackage{multirow}
\usepackage[section]{placeins}
\usepackage{array}
\usepackage{dcolumn}
\usepackage{booktabs}
\usepackage[detect-all]{siunitx}
\usepackage{makecell}
\usepackage{xcolor}
\usepackage{yfonts}
\usepackage[normalem]{ulem}
\usepackage{xcolor}
\usepackage[percent]{overpic}
\newcolumntype{P}[1]{>{\centering\arraybackslash}p{#1}}
\begin{document}

\title{Electronic Structure and Band-Edge Character of Ga-Substituted $\alpha$-Al$_2$O$_3$: A First-Principles Study}

\author{ Z. S. Machavariani$^{1}$, R. Ya. Kezerashvili$^{2,3,4}$, T. Tchelidze$^{1}$}
\affiliation{
$^{1}$
Ivane Javakhishvili Tbilisi State University, Tbilisi, Georgia\\
$^{2}$New York City College of Technology, The City University of New York,
Brooklyn, NY, USA\\
$^{3}$The Graduate School and University Center, The City University of New
York, New York, NY, USA\\
$^{4}$Long Island University, Brooklyn, NY, USA\\
}
\begin{abstract}
    
We present a first-principles density functional theory 
study of substitutional Ga incorporation in an $\alpha$-Al$_2$O$_3$-derived host, isolating chemical embedding effects without explicit geometric quantum confinement. 
Using a 30-atom $\alpha$-Al$_2$O$_3$ supercell with two substituted Ga atoms, we examine three nonequivalent Ga-pair configurations and combine configuration-dependent energetics, local structural analysis, Ga-centered distortion metrics, and electronic-structure calculations.
The energetically preferred configuration is governed not solely by Ga--O bond expansion but by the ability of the host lattice to accommodate bond-length and angular distortions around substituted Ga centers. Ga incorporation narrows the band gap relative to pristine $\alpha$-Al$_2$O$_3$ without generating mid-gap states, primarily modifying the conduction-band manifold while preserving the oxygen-dominated valence edge. Band-edge charge densities and charge-density-difference maps show that the electronic response remains localized around the Ga-centered coordination environment. These results establish a chemically resolved baseline for understanding Ga incorporation in ultrawide-band-gap oxides and for future studies of confinement and interface effects in Ga-containing oxide nanostructures.
\end{abstract}
\date{\today }
\maketitle

\section{Introduction}
\label{Intro}
Gallium oxide, Ga$_2$O$_3$, has emerged as a key ultrawide-band-gap oxide semiconductor for high-power electronics, deep-ultraviolet photodetectors, and related optoelectronic applications because of its large band gap, high breakdown field, and favorable materials platform for heterostructure engineering \cite{2,3,4,6,9,10}. In parallel, alloying and chemical mixing with other oxides such as Al$_2$O$_3$ have been explored as routes for modifying band edges, controlling band offsets, and tailoring local bonding environments \cite{7,8,23}.

Although Ga$_2$O$_3$ nanostructures and quantum-dot-like systems have attracted increasing attention \cite{11,12,13,14,15,16,17,18,19}, it is often difficult to separate pure chemical embedding effects from true geometric quantum confinement. Finite nanocrystals, explicit interfaces, dielectric contrast, and surface reconstruction can all modify the band gap and spatial character of frontier states. As a result, the intrinsic band-edge response produced solely by placing Ga in an Al–O host environment is not always easy to isolate.

Density functional theory (DFT) provides an effective atomistic framework for addressing this problem because it directly connects the local coordination environment to the orbital and spatial character of the electronic states \cite{20,21}. For a comprehensive review of first-principles point-defect calculations, including supercell and defect-thermodynamics methodology, see Ref. \cite{1}.  
Previous first-principles studies of Ga$_2$O$_3$, Al$_2$O$_3$, and (Al$_x$Ga$_1-x$)$_2$O$_3$ alloys have clarified phase stability, band-edge trends, optical properties, and defect-related behavior \cite{5,7,8,22,23,24}. However, a compact periodic model that isolates chemically embedded Ga-derived band-edge states inside an $\alpha$-Al$_2$O$_3$ host remains useful as a baseline for future explicit confinement studies.

In the present work, we therefore focus exclusively on a 30-atom $\alpha$-Al$_2$O$_3$-derived supercell and investigate the effect of replacing two Al atoms by Ga. This choice is intentional: the periodic supercell retains translational symmetry and suppresses envelope-state quantization characteristic of nanocryistal. Accordingly, changes in the band edges arise from local bonding, coordination, and chemical embedding rather than from finite-size confinement. The goal is not to model a fully confined Ga$_2$O$_3$ quantum dot, but to establish the structural and electronic baseline of a chemically embedded Ga–Al–O environment within $\alpha$-Al$_2$O$_3$.

The novelty of this work lies in combining configuration-dependent energetics, Ga-centered distortion metrics, and electronic-structure analysis to isolate the purely chemical embedding effects of substitutional Ga incorporation in an $\alpha$-Al$_2$O$_3$ host without explicit geometric quantum confinement. Unlike conventional alloy studies that primarily focus on band-gap trends, the present study  shows, that the energetic preference of Ga configurations is governed not simply by bond expansion, but by how the host lattice accommodates local bond-length and angular distortions around the Ga-centered coordination environment. Furthermore, Ga incorporation selectively restructures the conduction-band manifold to narrow the band gap without generating mid-gap states or disrupting the oxygen-dominated valence edge.

The study proceeds in three steps. First, we determine the relative stability of three nonequivalent two-Ga substitutional arrangements. Second, we quantify the local structural response through Ga–O bond-length and distortion analysis. Third, we analyze the electronic structure of the preferred configuration by combining DOS/PDOS calculations, real-space band-edge charge densities, host-versus-doped comparison, and charge-density-difference visualization. This strategy yields a consistent picture of how substitutional Ga perturbs the  $\alpha$-Al$_2$O$_3$ host at both the structural and electronic levels.

The paper is organized as follows. Sec. \ref{sec:computational_methods} presents the computational methods. The structural models, energetics, local relaxation, and distortion metrics are discussed in Sec. \ref{sec:structural_models}. Secs. \ref{TotalDensity} and \ref{HostComparison} analyze the electronic structure and host–doped comparison through DOS, PDOS, charge-density, and band-structure calculations. In Sec. \ref{ChargeDensity}, we examine the charge-density-difference maps associated with Ga incorporation. Finally, the main conclusions are summarized in Sec. \ref{Conclusion}, while Appendix \ref{app:distortion_metrics} collects the definitions of the Ga-centered distortion metrics. 

\section{Computational Methods}
\label{sec:computational_methods}

All calculations were performed within the framework of Kohn--Sham density functional theory using the plane-wave pseudopotential method as implemented in the Quantum ESPRESSO package \cite{25,29}. Exchange and correlation were treated within the generalized gradient approximation using the PBE functional \cite{26}. Projector augmented-wave (PAW) pseudopotentials were used for Al, Ga, and O; the canonical PAW formalism is described in Ref. \cite{27}.

The starting host structure was an $\alpha$-Al$_2$O$_3$-derived 30-atom supercell containing 12 cation sites and 18 oxygen atoms. The pristine host was fully relaxed at a plane-wave cutoff of 80 Ry and a charge-density cutoff of 640 Ry. Three nonequivalent two-Ga substitutional configurations, labeled A, B, and C, were then generated by replacing two Al atoms with Ga while keeping the relaxed $\alpha$-Al$_2$O$_3$ host cell fixed. These doped configurations were relaxed with respect to internal atomic coordinates using the same 80/640 Ry basis-set parameters.

For the final electronic-structure analysis, higher-accuracy single-point calculations were carried out at 90/720 Ry. Self-consistent field calculations used a $5\times5\times3$ Monkhorst--Pack mesh \cite{28}, and non-self-consistent DOS/PDOS calculations used an $8\times8\times6$ mesh with the tetrahedron method to produce smooth spectral distributions. Total DOS was obtained with \texttt{dos.x} and orbital-resolved PDOS with \texttt{projwfc.x}. The valence-band maximum (VBM) and conduction-band minimum (CBM) charge densities were extracted from the NSCF results using \texttt{pp.x} and analyzed from the corresponding cube files. Three-dimensional isosurfaces were visualized with VESTA \cite{30}.

A qualitative charge-density-difference map was constructed for the preferred configuration B as
$
\Delta \rho = \rho(\mathrm{B}) - \rho(\mathrm{host}),
$
by exporting total charge densities from \texttt{pp.x} to cube format and subtracting the pristine-host density from the doped density on compatible real-space grids. Because the pristine and doped systems do not contain the same total valence charge, this quantity is interpreted qualitatively as a spatial redistribution map rather than as a neutral deformation density or a rigorous charge-transfer metric.

The principal limitation of the present model arise from the use of semilocal PBE and from the fixed-cell supercell
construction. Semilocal PBE is known to underestimate the absolute band gaps of wide-gap oxides, so the present band-gap values are interpreted primarily in terms of relative trends, orbital character, and spatial redistribution rather than as quantitatively exact predictions of experimental optical gaps 
\cite{2,6,24,26}.  Additionally, employing a fixed-cell relaxation strategy neglects full lattice volume and shape adjustments, while the 30-atom supercell restricts the thermodynamic analysis to relative energy differences rather than full defect formation energies. These computational choices represent inherent model limitations that restrict the study primarily to qualitative and relative trends in structural distortion and electronic state redistribution, rather than quantitative predictions of absolute experimental band gaps. 
\section{Structural Models, Energetics, and Local Relaxation}
\label{sec:structural_models}

\subsection{Nonequivalent Two-Ga Configurations}
\label{subsec:two_ga_configurations}

To examine how the relative placement of two substitutional Ga atoms influences the stability of the $\alpha$-Al$_2$O$_3$ host, three nonequivalent configurations were constructed in the same 30-atom supercell. Their initial Ga--Ga separations span compact, intermediate, and far-separated arrangements, allowing the energetic role of local geometry to be assessed systematically.

\begin{figure}[!htbp]
    \centering
    \includegraphics[width=0.75\textwidth]{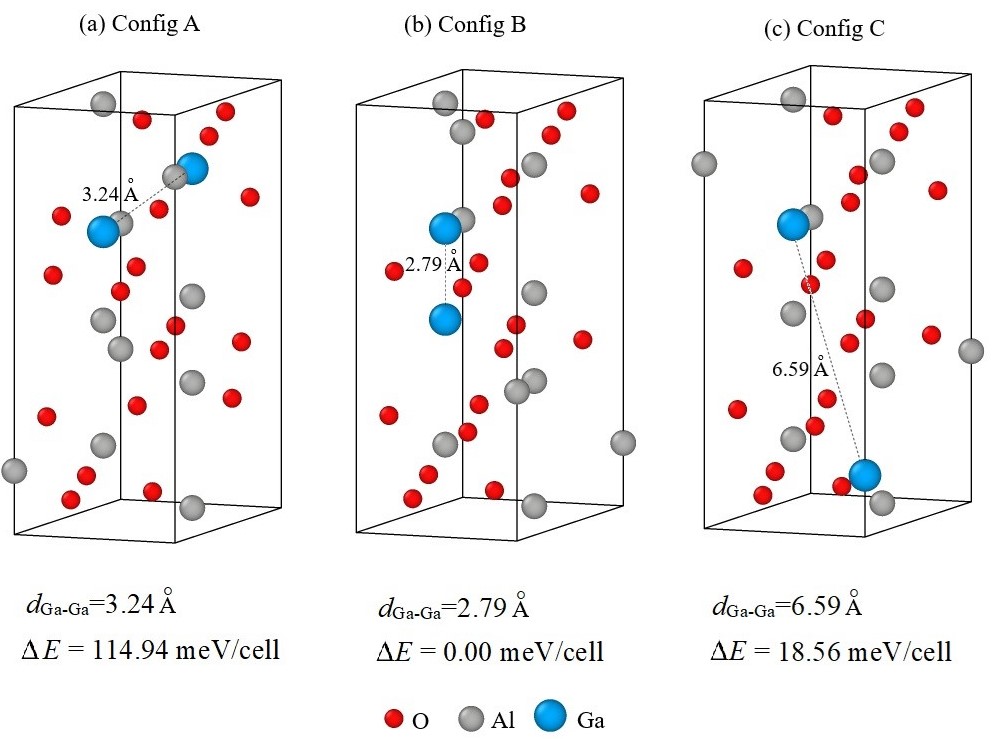}
    \caption{Relaxed two-Ga substitutional configurations in the 30-atom $\alpha$-Al$_2$O$_3$-derived supercell: (a) configuration A, (b) configuration B, and (c) configuration C. Oxygen, aluminum, and gallium atoms are shown in red, gray, and blue, respectively. The dashed line indicates the relaxed Ga--Ga separation in each configuration.}
    \label{fig:ga_configurations}
\end{figure}

Figure~\ref{fig:ga_configurations} shows that the three structures differ only by the relative placement of the substituted Ga pair, yet this apparently modest geometric change produces clear energetic differences after relaxation. Configuration B is the lowest-energy arrangement, configuration C is metastable but close in energy, and configuration A is substantially less favorable. The preferred structure is therefore not the most spatially separated pair but a compact local Ga arrangement, indicating that the host lattice can accommodate certain nearby Ga placements more effectively than others.

\begin{table}[!htbp]
	\centering
	\fontsize{9}{11}\selectfont
	\setlength{\tabcolsep}{2pt}
	\renewcommand{\arraystretch}{1.12}
	
	\caption{Structural and energetic summary of the three relaxed two-Ga substitutional configurations in $\alpha$-Al$_2$O$_3$.}
	
	\begin{tabular}{@{}c c c c c c c c@{}}
		\hline
		
		Config &
		\begin{tabular}{c}
			Initial\\
			Ga--Ga\\
			(\AA)
		\end{tabular} &
		\begin{tabular}{c}
			Relaxed\\
			Ga--Ga\\
			(\AA)
		\end{tabular} &
		\begin{tabular}{c}
			Final relax\\
			energy\\
			(Ry)
		\end{tabular} &
		\begin{tabular}{c}
			$\Delta E$ vs B\\
			(meV/cell)
		\end{tabular} &
		\begin{tabular}{c}
			Final total\\
			force\\
			(Ry/Bohr)
		\end{tabular} &
		\begin{tabular}{c}
			Max atomic\\
			force\\
			(Ry/Bohr)
		\end{tabular} &
		\begin{tabular}{c}
			Final\\
			pressure\\
			(kbar)
		\end{tabular}
		\\
		
		\hline
		
		A &
		3.2552 &
		3.2397 &
		-1704.41622669 &
		114.94 &
		0.000280 &
		$8.83\times10^{-5}$ &
		58.66 \\
		
		B &
		2.6815 &
		2.7918 &
		-1704.42467465 &
		0.00 &
		0.000202 &
		$6.27\times10^{-5}$ &
		56.90 \\
		
		C &
		6.6785 &
		6.5862 &
		-1704.42331023 &
		18.56 &
		0.000234 &
		$8.29\times10^{-5}$ &
		57.09 \\
		
		\hline
	\end{tabular}
	
	\label{tab:configuration_energies}
	
\end{table}

Table~\ref{tab:configuration_energies} confirms the energetic ordering $
\mathrm{B} < \mathrm{C} \ll \mathrm{A}.
$
Configuration C lies only 18.56 meV/cell above B, whereas configuration A is 114.94 meV/cell higher. This shows that stability is not determined by the Ga--Ga distance alone. The compact B arrangement is energetically preferred, implying that local site geometry and the response of the host lattice are more important than simple pair separation. Because the doped structures were relaxed at fixed lattice
vectors, the nonzero residual pressures reported in Table \ref{tab:configuration_energies} reflect the imposed host-cell constraint and should not be
interpreted as equilibrium hydrostatic pressures

A full thermodynamic treatment would require defect formation energies referenced to appropriate Al, Ga, and O chemical potentials, together with finite-temperature free-energy contributions and configurational statistics \cite{1}. Such an analysis is outside the scope of the present study. Within the fixed-cell model and among the three configurations examined, configuration B is the lowest-energy arrangement.

The relatively small energy difference between configurations B and C, 18.56 meV/cell, is comparable to the thermal-energy scale at room temperature, suggesting that configuration C may also be thermally accessible. In contrast, configuration A, which lies 114.94 meV/cell above B, is substantially less favorable. The resulting energetic ordering therefore indicates that local Ga-pair stability is governed by the interplay between configurational energetics, local structural distortion, and the ability of the $\alpha$-Al$_2$O$_3$ host lattice to accommodate the substituted Ga atoms.

\subsection{Local Ga--O Bond-Length Response}

To connect the energetic ordering with local structural relaxation, we analyzed the six nearest O neighbors around each substituted Ga site and compared these bond lengths with the corresponding Al–O environment at the same crystallographic sites in the pristine host.

\begin{table}[!htbp]
	\centering
	\fontsize{9}{11}\selectfont
	\setlength{\tabcolsep}{4pt}
	\renewcommand{\arraystretch}{1.15}
		\caption{Local Ga--O bond-length characteristics for the three relaxed two-Ga substitutional configurations in $\alpha$-Al$_2$O$_3$.}
	
	\begin{tabular}{c c c c c c c}
		\hline \hline
		
		Config &
		\begin{tabular}{c}
			Ga site\\
			indices\\
			~
		\end{tabular} &
		\begin{tabular}{c}
			Average\\
			Ga--O\\
			(\AA)
		\end{tabular} &
		\begin{tabular}{c}
			Min\\
			Ga--O\\
			(\AA)
		\end{tabular} &
		\begin{tabular}{c}
			Max\\
			Ga--O\\
			(\AA)
		\end{tabular} &
		\begin{tabular}{c}
			Host reference\\
			average Al--O\\
			(\AA)
		\end{tabular} &
		\begin{tabular}{c}
			Average expansion\\
			vs host\\
			(\AA)
		\end{tabular}
		\\
		
		\hline
		
		A & 1, 2 & 1.9811 & 1.9007 & 2.0615 & 1.9340 & $+0.0471$ \\
		B & 2, 9 & 1.9956 & 1.9239 & 2.0673 & 1.9340 & $+0.0616$ \\
		C & 2, 5 & 1.9930 & 1.9193 & 2.0668 & 1.9340 & $+0.0590$ \\
		
		\hline
	\end{tabular}
	
	\label{tab:Table2}
	
\end{table}

Table \ref{tab:Table2} shows that Ga substitution consistently expands the local cation–oxygen coordination shell relative to the pristine-host reference value of $1.9340~\text{\AA}$. The average Ga–O bond lengths range from $1.9811~\text{\AA}$ to $1.9956~\text{\AA}$, corresponding to local expansions of +0.0471 to $+0.0616~\text{\AA}$. The largest average expansion occurs in configuration B, which nevertheless remains the lowest-energy structure. This observation is important because it shows that energetic preference is not determined simply by minimizing local bond elongation. Instead, the preferred configuration emerges from a balance between local Ga–O relaxation, the relative placement of the Ga pair, and the overall elastic response of the $\alpha$-Al$_2$O$_3$ host.

\subsection{Ga-Centered Distortion Metrics}
\label{Distortion_Metrics}

To quantify local geometric irregularity beyond average Ga–O bond expansion, several Ga-centered distortion metrics were evaluated from the six nearest oxygen neighbors around each substituted Ga site. These metrics distinguish average bond-length expansion from bond-length disorder and angular distortion. Formal definitions of the metrics and their notation are collected in Appendix A.

\begin{table}[!htbp]
	\centering
	\caption{Ga-centered distortion metrics for the three relaxed two-Ga substitutional configurations in $\alpha$-Al$_2$O$_3$. The pristine-host reference values correspond to the same crystallographic sites before Ga substitution.}
	\label{tab:distortion_metrics}
	
	\fontsize{9}{11}\selectfont
	\setlength{\tabcolsep}{8pt}
	\renewcommand{\arraystretch}{1.15}
	
	\begin{tabular}{c c c c c c}
		\hline \hline
		System &
        Ga site indices &
        $D$ &
        \begin{tabular}{c}
            $\sigma$\\
            (\AA)
        \end{tabular} &
        $\lambda$ &
        \begin{tabular}{c}
            $\sigma_{\theta}^{2}$\\
            $(^\circ)^2$
        \end{tabular} \\
		\hline
		
		Host ref. & same sites & 0.0299 & 0.0578 & 1.000894 & 98.882 \\
		A         & 1,2        & 0.0406 & 0.0804 & 1.001649 & 126.953 \\
		B         & 2,9        & 0.0359 & 0.0717 & 1.001291 & 116.836 \\
		C         & 2,5        & 0.0370 & 0.0737 & 1.001368 & 141.310 \\
		
		\hline
	\end{tabular}
	\label{Table3}
\end{table}

Table~\ref{tab:distortion_metrics} shows that the host reference octahedra at the corresponding Al sites are more regular than the Ga-centered shells in all three doped configurations. Relative to the host reference, the bond distortion index \textit{D} and the bond-length standard deviation $\sigma$ increase in every case, while the angle variance $\sigma_\theta^2$ also increases substantially. This confirms that Ga substitution produces not only bond expansion, as already seen in Table~\ref{tab:Table2}, but also measurable bond-length disorder and angular distortion.

The distortion metrics refine the structural interpretation obtained from the bond-length averages alone. All three Ga-substituted configurations are more distorted than the pristine-host reference, both in bond-length distribution and in angular regularity. Configuration A exhibits the largest bond-length distortion, configuration C the strongest angular distortion, and configuration B the smallest bond-length distortion among the three doped structures. Thus, the preferred structure is not the one with the smallest average Ga–O expansion, but rather the one in which the expanded coordination shell is accommodated with comparatively lower internal distortion. These results provide a more precise structural explanation for why configuration B is energetically preferred.

\section{Electronic Structure of the Preferred Configuration B}
\label{TotalDensity}
\subsection{Total and Projected Density of States}

\begin{figure}[!htbp]
    \centering
    \includegraphics[width=0.93\textwidth]{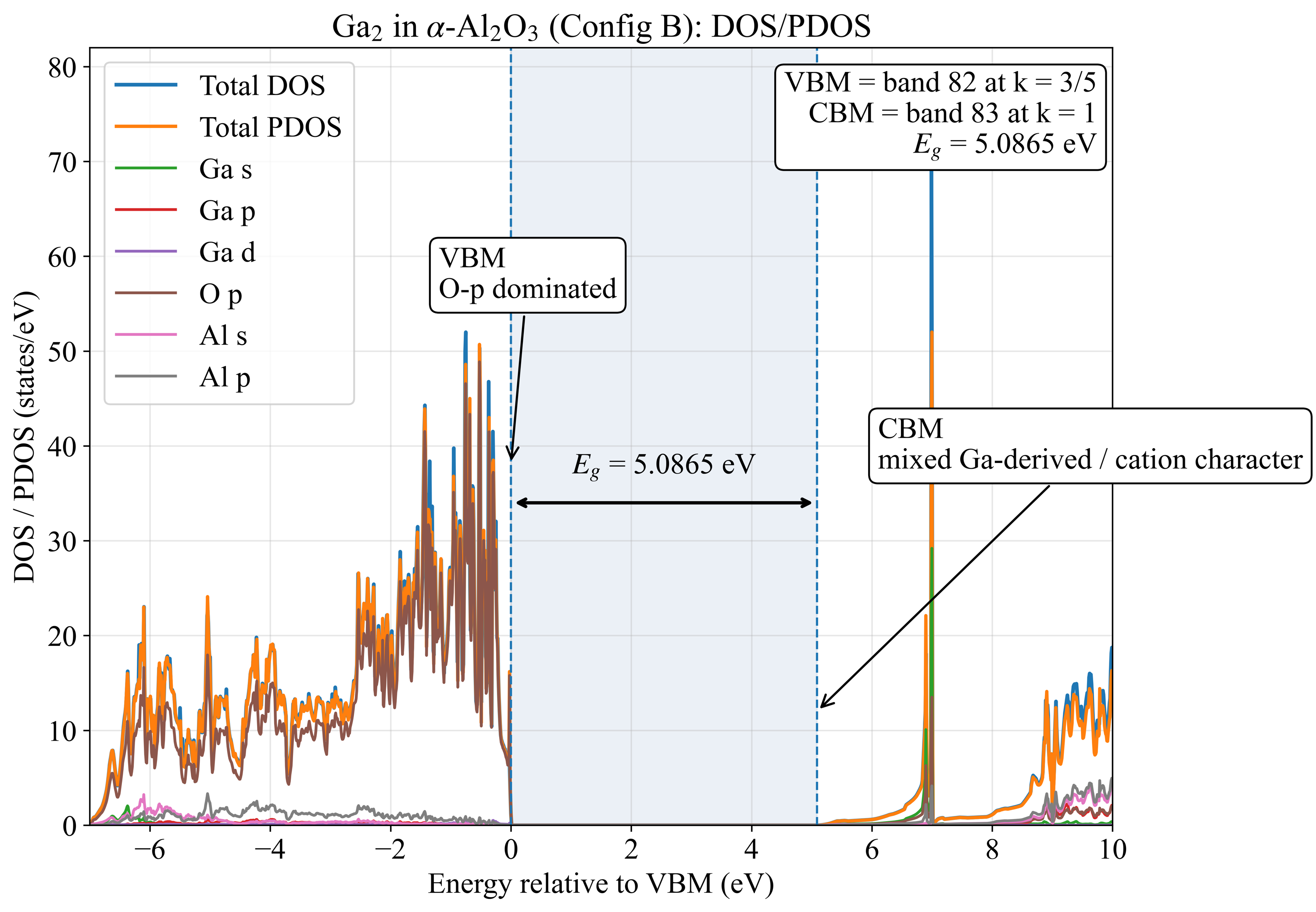}
    \caption{Total DOS and orbital-resolved PDOS for the preferred two-Ga configuration B in $\alpha$-Al$_2$O$_3$. The energy axis is referenced to the VBM. The VBM corresponds to band 82 at $k=3/5$ and the CBM to band 83 at $k=1$, yielding an indirect band gap of $5.0865~\mathrm{eV}$.}
    \label{fig:Fig2}
\end{figure}

Figure \ref{fig:Fig2} shows that configuration B remains an insulating system with an indirect Kohn–Sham band gap of 5.0865 eV. The upper valence band is dominated by O-p states, indicating that the valence edge preserves the characteristic anionic framework of the oxide host. Near the VBM, Ga- and Al-derived contributions remain comparatively small.

The conduction-band onset is modified more strongly. The lowest conduction states contain substantial Ga-derived and other cationic contributions, showing that Ga incorporation primarily perturbs the conduction manifold rather than the valence manifold. Deep Ga-d peaks appear well below the valence edge and do not control the fundamental gap.

\subsection{Real-Space Band-Edge Charge Densities}

\begin{figure}[!htbp]
    \centering
    \includegraphics[width=0.93\textwidth]{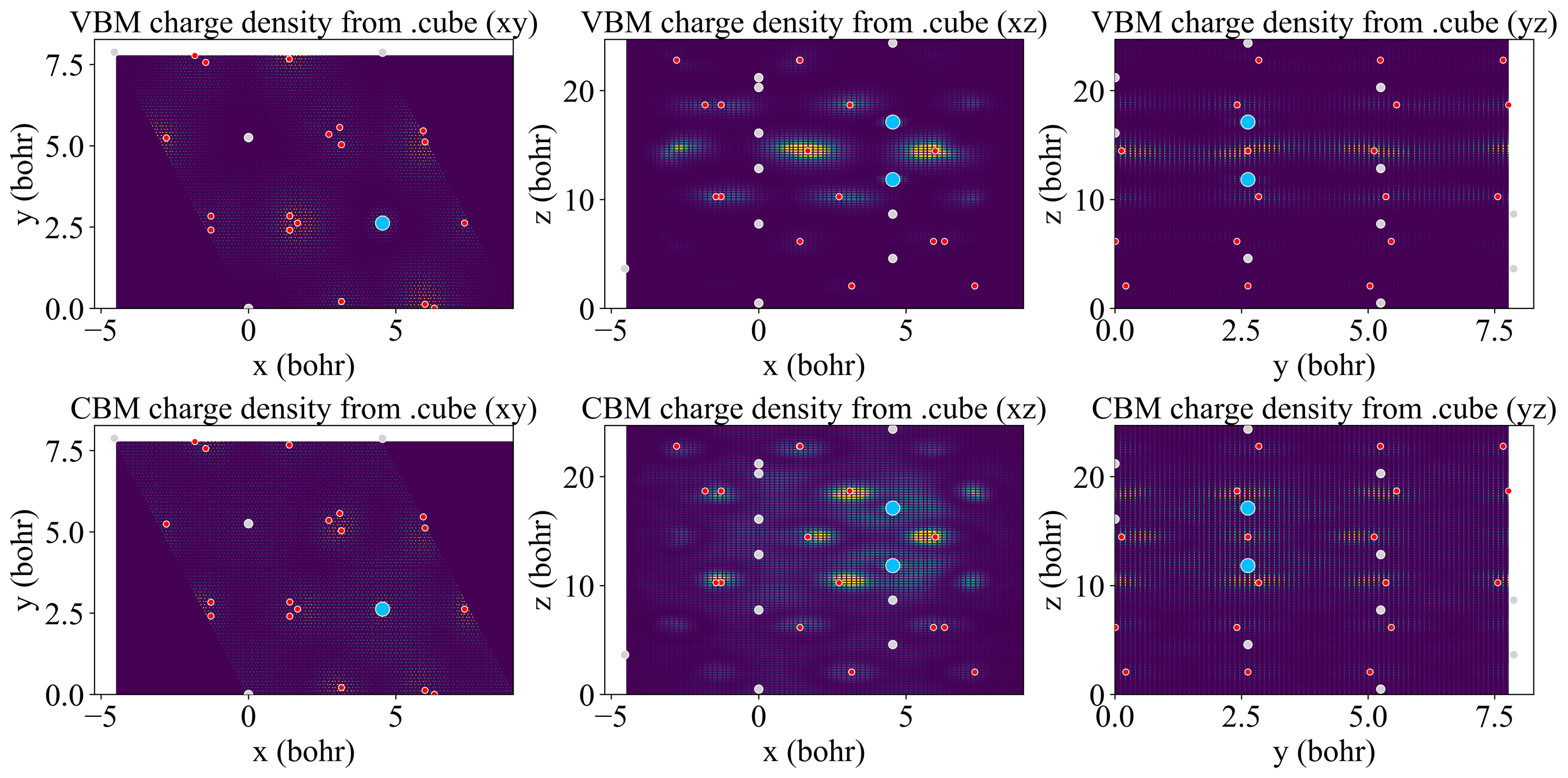}
    \caption{Real-space charge-density projections for the VBM and CBM of configuration B, obtained from the corresponding cube files. For each state, three two-dimensional projections are shown ($xy$, $xz$, and $yz$).}
    \label{fig:Fig3}
\end{figure}

Figure~\ref{fig:Fig3} shows that the VBM charge density is concentrated mainly on oxygen sites, consistent with the O-p-dominated valence edge identified in the DOS/PDOS analysis. By contrast, the CBM charge density is more spatially extended and distributed over the local Ga–O coordination environment and neighboring cation network.

This real-space picture shows that the conduction-band minimum is not a purely Ga-localized impurity state. Instead, it is a mixed Ga–O–cation state whose character emerges from the locally distorted coordination environment created by Ga substitution.

\section{Host–Doped Comparison}
\label{HostComparison}
\subsection{Total DOS/PDOS Comparison}

\begin{figure}[!htbp]
    \centering
    \includegraphics[width=0.93\textwidth]{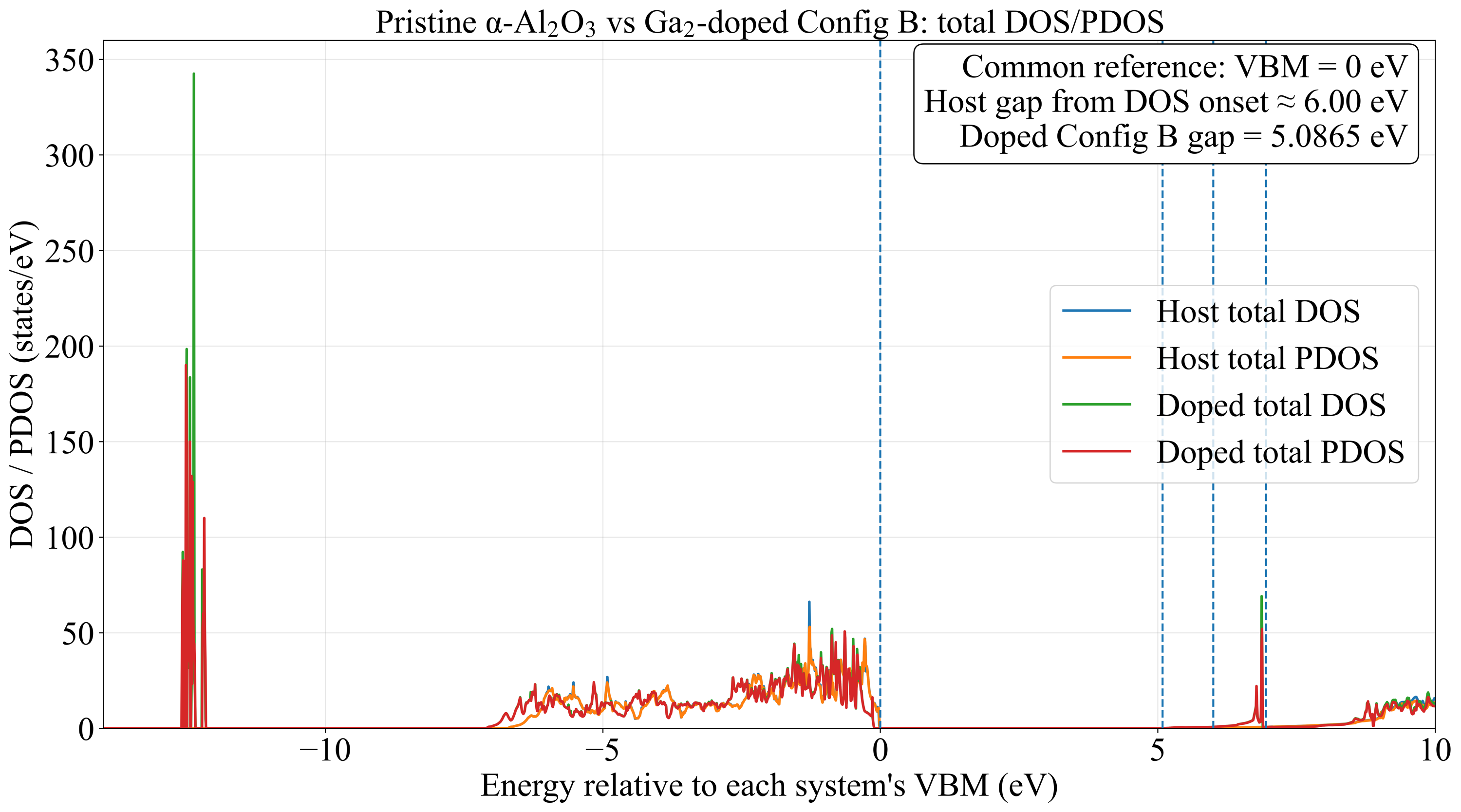}
    
    \caption{Comparison of the total DOS and total PDOS of pristine $\alpha$-Al$_2$O$_3$ and Ga$_2$-doped configuration B. For both systems, the energy scale is shifted so that the VBM is set to $0~\mathrm{eV}$.}
    \label{fig:Fig4}
\end{figure}

Figure~\ref{fig:Fig4} compares pristine $\alpha$-Al$_2$O$_3$ with configuration B and shows that Ga substitution narrows the gap from about 6.00 eV (estimated from the host DOS onset) to 5.0865 eV for configuration B. Importantly, this reduction is not accompanied by the appearance of mid-gap states.
The valence-band region of the host and doped systems remains broadly similar, which indicates that the oxygen-derived valence manifold is comparatively robust against Ga substitution. In contrast, the conduction-band onset in the doped structure shifts downward and gains additional low-energy spectral weight, confirming that the main electronic consequence of Ga incorporation is a restructuring of the conduction manifold.

\subsection{Orbital Origin of the Host–Doped Difference}

\begin{figure}[!htbp]
    \centering
    \includegraphics[width=0.93\textwidth]{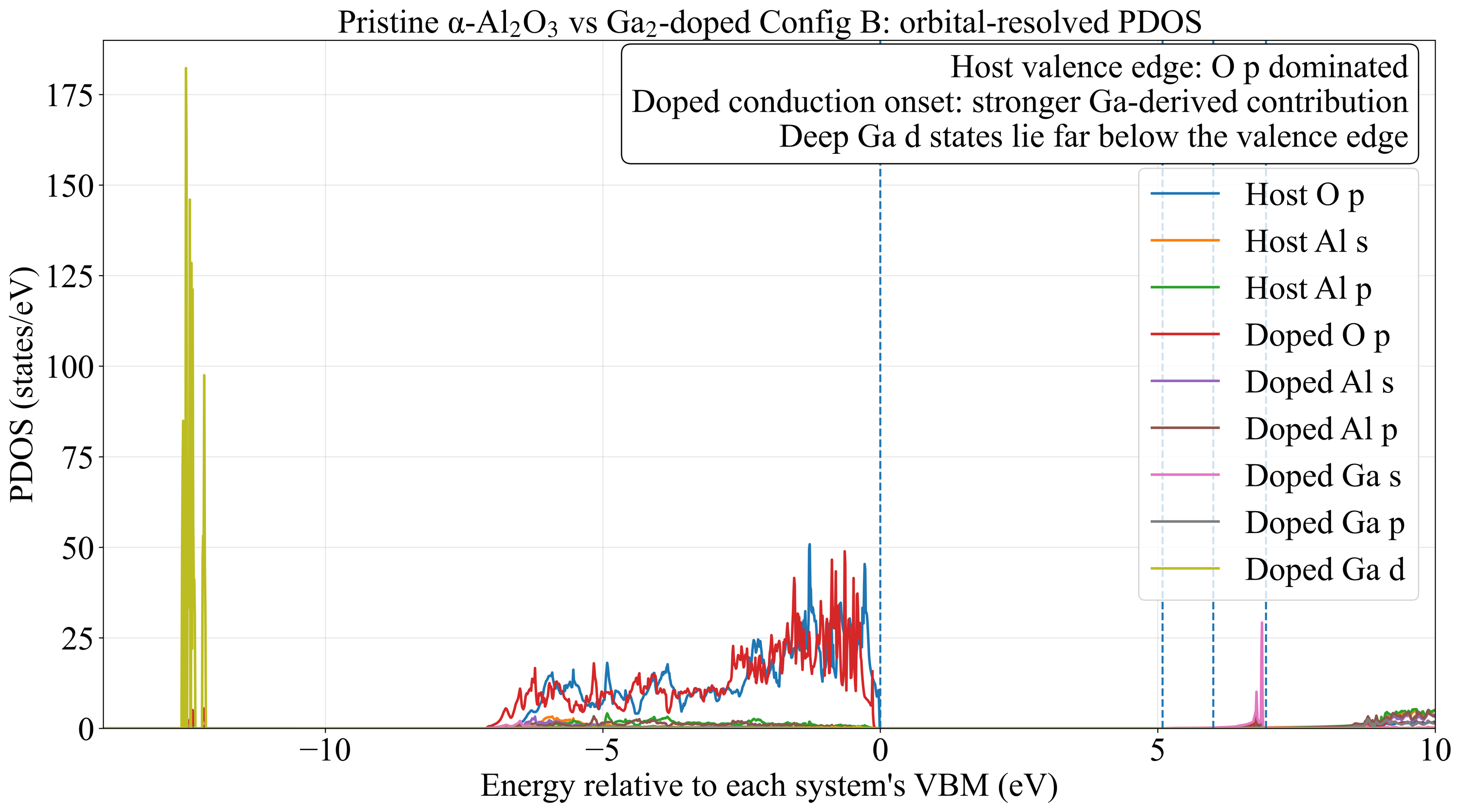}
    
    \caption{Orbital-resolved PDOS for pristine $\alpha$-Al$_2$O$_3$ and Ga$_2$-doped configuration B on a common VBM-referenced energy scale.}
    \label{fig:Fig5}
\end{figure}

Figure~\ref{fig:Fig5} clarifies the orbital origin of the host–doped difference. In pristine $\alpha$-Al$_2$O$_3$, the upper valence region is dominated by O-p states, whereas Al-derived states remain minor near the band edge. In the doped system, this qualitative valence-band picture is preserved: the VBM remains predominantly O-p-like.
The main change appears on the conduction side, where Ga-derived contributions become significant near the band onset. Deep Ga-d states remain far below the valence edge and do not determine the gap. Taken together, Figures~\ref{fig:Fig4} and~\ref{fig:Fig5} show that substitutional Ga primarily perturbs the conduction manifold while preserving the host-like oxygen-dominated valence edge.

\subsection{Site-Resolved Local PDOS Comparison}

\begin{figure}[!htbp]
    \centering
    \includegraphics[width=0.93\textwidth]{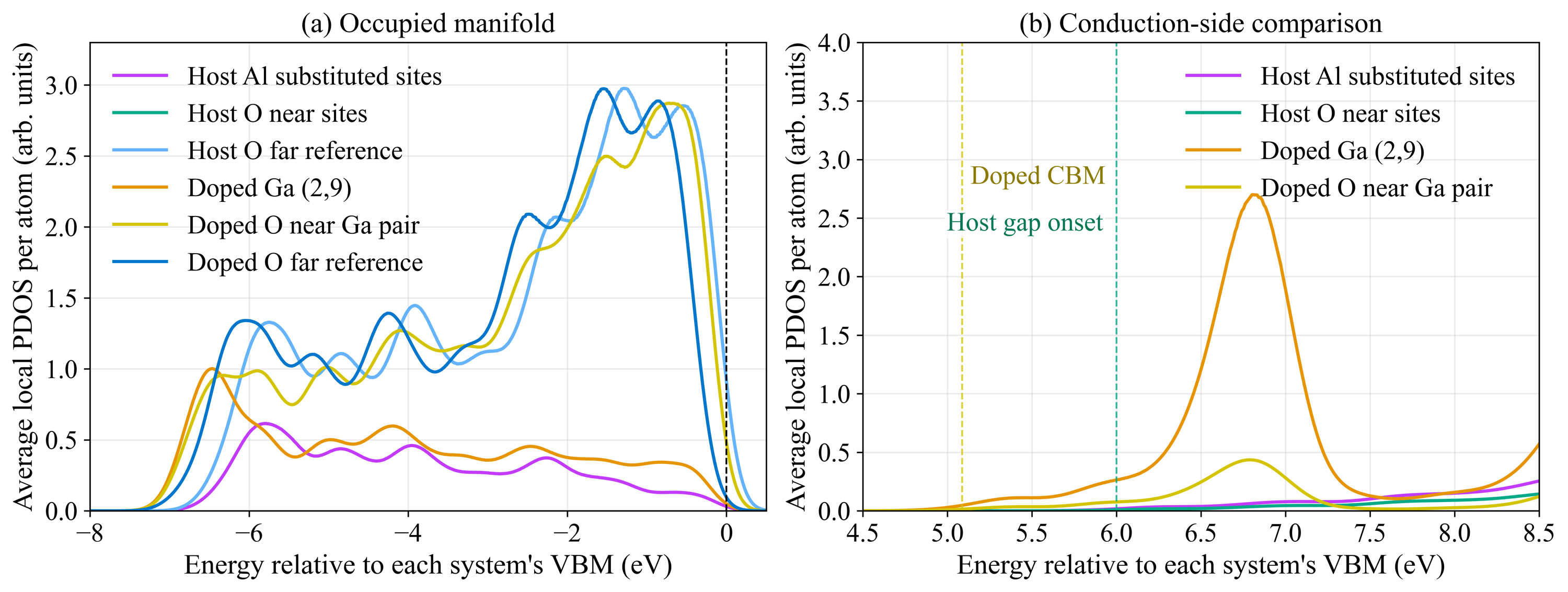}
    
    \caption{Normalized site-resolved local projected density of states (PDOS) for pristine $\alpha$-Al$_2$O$_3$ and $\alpha$-Ga$_2$-doped configuration B, plotted as average local PDOS per atom for each selected group and aligned to the valence-band maximum of each system. (a) Occupied manifold. (b) Conduction-side comparison in the energy range containing the host and doped conduction-band onsets. The dashed vertical line at 0 eV marks the VBM in each panel, while the colored dashed lines in panel (b) indicate the doped CBM and the approximate host gap onset.}
    \label{fig:Fig6}
\end{figure}

Figure~\ref{fig:Fig6} provides a localized view of the host–doped electronic response. In the occupied manifold, the dominant spectral weight remains associated with oxygen in both pristine $\alpha$-Al$_2$O$_3$ and the Ga$_2$-doped structure, confirming that the valence-band region preserves predominantly anionic character. By contrast, the conduction-side comparison shows that Ga substitution introduces pronounced low-lying unoccupied spectral weight within the host-gap region, with the strongest enhancement occurring on the Ga sites and a smaller but visible contribution from the neighboring oxygen shell. The largest Ga-centered feature lies somewhat above the conduction-band onset, implying that the CBM is not a purely on-site Ga impurity level but a mixed and more spatially distributed conduction-edge state.

The local-PDOS analysis therefore supports the broader picture obtained from the global DOS/PDOS, band-structure, and charge-density visualizations: substitutional Ga primarily perturbs the conduction manifold, whereas the oxygen-dominated valence framework of the $\alpha$-Al$_2$O$_3$ host remains comparatively robust.

\subsection{Comparative Band-Structure Analysis}

\begin{figure}[!htbp]
    \centering
    \includegraphics[width=0.93\textwidth]{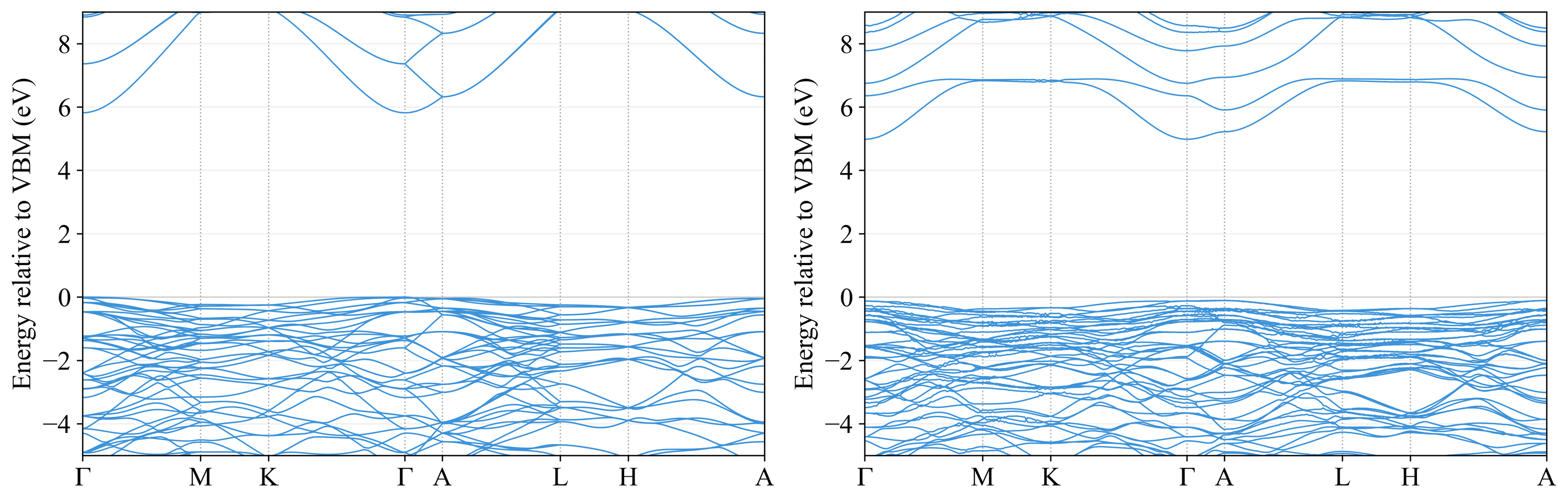}
    \caption{Comparative band structures of pristine $\alpha$-Al$_2$O$_3$ (left panel) and the preferred Ga$_2$-doped configuration B (right panel) along the $\Gamma$–M–K–$\Gamma$–A–L–H–A path, with the valence-band maximum set to 0 eV in each panel. The pristine host exhibits a wider insulating gap of about 6.0 eV, whereas configuration B remains insulating but shows a reduced indirect gap near 5.0 eV. The doped dispersion is somewhat less smooth because of symmetry reduction and more difficult empty-state convergence in the substituted supercell, but the near-gap region is sufficiently clear for qualitative host–doped comparison.}
    \label{fig:Fig7}
\end{figure}

Figure~\ref{fig:Fig7} provides direct reciprocal-space confirmation of the host–doped comparison inferred earlier from the DOS and PDOS analysis. Pristine $\alpha$-Al$_2$O$_3$ remains a wide-gap insulator, while the preferred Ga$_2$-doped configuration B also remains insulating but exhibits a smaller gap. The comparison shows that the dominant electronic change occurs on the conduction side, where the lowest unoccupied bands of the doped system are shifted downward relative to the host. By contrast, the upper valence manifold is modified more modestly, consistent with the conclusion that the oxygen-derived valence edge of the host is comparatively robust against substitutional Ga incorporation.

The band-structure comparison therefore supports the same picture obtained from the DOS/PDOS and real-space charge-density analysis: Ga substitution narrows the gap primarily through a restructuring of the conduction manifold rather than through the formation of mid-gap defect states or a major reconstruction of the valence edge. The doped dispersion is visibly rougher than the pristine-host reference because the substituted supercell has lower symmetry and is numerically more difficult to converge, so the figure is used here as qualitative supporting evidence rather than as a basis for precision analysis of band curvature or effective masses. Nevertheless, the near-gap region is sufficiently clear to confirm that configuration B preserves the insulating character of the $\alpha$-Al$_2$O$_3$ matrix while lowering the conduction-band edge.

\section{Charge-Density-Difference Analysis}
\label{ChargeDensity}

\begin{figure}[!htbp]
	\centering
	\includegraphics[height=0.45\textheight,keepaspectratio]{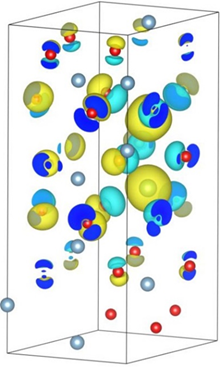}
	
	\caption{Charge-density-difference isosurfaces for configuration B, defined as $\Delta \rho = \rho(B) - \rho(\mathrm{host})$. Positive and negative isosurfaces are shown at $\pm 0.03$ in arbitrary electron-density units.}
	\label{fig:Fig8}
\end{figure}

Figure~\ref{fig:Fig8} shows that the electronic response to Ga substitution is strongly localized in the vicinity of the substituted Ga pair and the surrounding oxygen coordination environment. Both charge accumulation and charge depletion appear predominantly around the local Ga-centered bonding network rather than being distributed uniformly throughout the supercell.

This behavior is consistent with the structural results, which showed systematic Ga–O bond expansion, and with the DOS/PDOS analysis, which indicated that Ga incorporation primarily modifies the conduction-side electronic structure. Because $\Delta \rho = \rho(B) - \rho(\mathrm{host})$ is constructed from two systems with different total valence charge, it is interpreted here qualitatively as a visualization of spatial redistribution rather than as a quantitative deformation-density or charge-transfer analysis.

\section{Conclusions}
\label{Conclusion}

We have presented a first-principles DFT study of substitutional Ga incorporation in an $\alpha$-Al$_2$O$_3$-derived host, with the explicit aim of isolating chemical embedding effects from geometric quantum confinement. Three nonequivalent two-Ga substitutional arrangements were examined, and configuration B was identified as the energetically preferred structure.
Local bond-length analysis showed that Ga substitution expands the cation–oxygen coordination shell in all three configurations, with the largest average Ga–O expansion occurring in the lowest-energy structure. Ga-centered distortion metrics further show that the doped coordination shells are more distorted than the pristine-host reference in both bond lengths and bond angles. Among the three doped structures, configuration A exhibits the largest bond-length distortion, configuration C the strongest angular distortion, and configuration B the smallest bond-length distortion despite having the largest average Ga–O expansion. This demonstrates that energetic preference is not determined solely by minimizing local expansion or distortion; instead, it reflects the interplay between local coordination geometry, pair arrangement, and host-lattice response.
The detailed electronic-structure analysis of configuration B yielded an indirect Kohn–Sham band gap of 5.0865 eV. The valence-band edge remains oxygen dominated, whereas the conduction-band onset acquires strong Ga-derived and broader cationic character. Host–doped comparison shows that Ga substitution narrows the gap relative to pristine $\alpha$-Al$_2$O$_3$ without creating mid-gap states.
Real-space VBM/CBM charge densities and the qualitative charge-density-difference map confirm that the electronic response is strongly localized in the Ga-centered coordination environment. The results therefore support a chemically embedded picture in which substitutional Ga perturbs primarily the conduction manifold while preserving the host-like oxygen-derived valence framework. This chemically resolved baseline is an appropriate starting point for future studies that will introduce explicit finite-size confinement, dielectric contrast, and interface reconstruction in Ga-containing ultrawide-band-gap oxide nanostructures.

\appendix

\section{Definitions of Ga-Centered Distortion Metrics}
\label{app:distortion_metrics}

The formulas used for the Ga-centered distortion analysis in Subsec.~\ref{Distortion_Metrics} are collected here for completeness.

\[
\sigma =
\left[
\frac{1}{6}
\sum_{i=1}^{6}
\left(d_i-\bar{d}\right)^2
\right]^{1/2},
\qquad
D =
\frac{1}{6}
\sum_{i=1}^{6}
\frac{\left|d_i-\bar{d}\right|}{\bar{d}},
\]

\[
\lambda =
\frac{1}{6}
\sum_{i=1}^{6}
\left(\frac{d_i}{\bar{d}}\right)^2,
\qquad
\sigma_{\theta}^{2} =
\frac{1}{15}
\sum_{j=1}^{15}
\left(
\theta_j-\theta_j^{\mathrm{ideal}}
\right)^2,
\]
where $d_i$ ($i=1,\ldots,6$) are six nearest Ga--O distances,
$\bar{d}$ is the average of the six nearest Ga--O distances,
$D$ is the bond distortion index,
$\sigma$ is the bond-length standard deviation,
$\lambda$ is the quadratic elongation relative to the average bond length,
and $\sigma_{\theta}^{2}$ is the mean squared deviation of the fifteen
O--Ga--O angles from the nearest ideal octahedral value, $90^\circ$ or
$180^\circ$, and $\theta_j^{\mathrm{ideal}}$ is chosen as whichever of
$90^\circ$ or $180^\circ$ is closer to the actual angle.

Because the Ga-centered coordination shell is defined by six nearest oxygen
neighbors, there are 
$\binom{6}{2}=15$
distinct O--Ga--O pair angles; the angle variance is, therefore, evaluated as
the mean squared deviation over these 15 angles.

\section*{Acknowledgments}

The authors are grateful to Dr. George Jackeli (Max Planck Institute for Solid State Research and Institute for Functional Matter and Quantum Technologies, University of Stuttgart) for fruitful discussions, valuable comments, and helpful advice.

\end{document}